\documentstyle[11pt,newpasp,twoside,epsf]{article}
\def\edcomment#1{\iffalse\marginpar{\raggedright\sl#1\/}\else\relax\fi}
\reversemarginpar

\begin{document}
\title{Professional Astronomy without a Librarian}
\author{Heinz Andernach}
\affil{Depto.\ de Astronom\'{\i}a, Univ.\ de Guanajuato, Apdo.\ Postal 144,
Guanajuato, C.P.\ 36000, Mexico}

\begin{abstract}
Virtually every ``serious'' place where professional astronomy 
is done has a librarian, even if shared with the physics or math department.
Since its creation in 1994 of {\it Departamento de Astronom\'{\i}a} (DA) of
Universidad de Guanajuato (UG) it was neither provided with a librarian,
nor with proper space for its holdings, nor with a budget allowing 
institutional journal subscriptions.  
I~describe my experience of now five years as ``amateur'' 
librarian, and present information on other small astronomy
institutions in Mexico in a similar situation.
\end{abstract}

\section{The Place}
Since 1997 the DA of UG (DAUG) is housed at 2200\,m above sea level,
overlooking the city of Guanajuato, close to the ``geometrical
center'' of Mexico, and declared as ``heritage of mankind'' by UNESCO.
The research staff of eight \linebreak[4]
astronomers participates in
undergraduate teaching within the Physics and other programs at UG,
and hopes to offer a postgraduate program in astrophysics soon.
The DA maintains the {\it Observatorio La Luz}
with a 57-cm optical reflector, used for public outreach purposes and 
being prepared as a student laboratory.

\section{Origins of our Library}
The two founder-members of the DA had already collected (by donations)
a few decades of {\it ApJ, AJ}, and {\it MNRAS}, but other major journals
(like {\it A\&A}) had been subscribed only since 1994.
Since my arrival, in late 1996, I volunteered as ``provisional''
librarian, given the lack of a professional Department librarian
hired by UG. The central library of UG was too far away to manage
efficiently the library of the DA.
As a matter of fact this situation continues until now. 
 
After preparing the first inventory of our holdings in early
1997, I posted an inquiry to ASTROLIB for donations to fill 
the holes of our journal coverage.  This caused a wave of offers
from professional librarians all over the world.  Hundreds of kilos of
journals were received over the following months.  In June 1997 the DA
moved to its present building with only 6 offices for 10 people,
and no library~room. However, our neighbor, the Maths Research Center 
``CIMAT'', \linebreak[4] maintained by the Mexican science foundation CONACyT,
generously offered $\sim$\,60\,m$^2$ of their library ``provisionally''
for the DA. CIMAT is $\sim$\,200\,m away ($\sim$\,40\,m
vertical\,!), {\it but\,} it {\it closes} during nights and weekends which 
never made it attractive for a ``leisure visit''. After five years
nothing has changed\,!  

\section{Our Present Library}
As it was not practical to store all our holdings ``up there''
at CIMAT, we decided to use all possibly available space
in the offices and corridors for the most recent journals
($\sim$\,1 year back) and modern books (for teaching and research).
Until now our hopes to obtain a dedicated library were not fulfilled,
despite a written promise and money allocation by UG in spring 2001.
Thus the journals have grown hopelessly beyond their initially assigned
growth space. Maintaining the alphabetic order of journals would now
imply a full rearrangement of our holdings which is beyond our 
available manpower.  Moreover, in the meantime most of our journal
holdings have become available freely from ADS, making a visit to the
physical library each time less attractive to our staff.
However, downloading of
articles is often limited to weekends given our slow internet line.

In the absence of a librarian, or other suitable personnel, I~only
seal the books with a DA stamp but do not afford to assign them 
a catalog number, so they have no ``reproducible'' shelf location yet.
Moreover, in mid-1999 a heavy rain and a leaky window affected
part of our shelves at CIMAT, and about 15\% of our book holdings, many of
them of historic interest, were waterlogged.  Most of the book holdings
had to be removed in a hurry (in my absence) and since then only very
limited ``order'' was re-established.  I visit the library at most once
a month to accommodate what does not fit any more in our office building.

\section{Subscriptions}
Since its foundation the DA received the six major astronomy journals
by personal subscription. The idea was that we would
gradually be able to pay institutional rates, but with a
2002 budget of USD\,3000 we still cannot afford a single
journal at the institutional rate. Thanks to the generous permit
by some publishers to continue with the personal rate,
we manage to subscribe to
{\it ApJ, ApJS, AJ, MNRAS, A\&A, Nature, PASA, PASJ, PASP,
BAAS, S\&T}, and {\it Mercury}.
These regular funds also allow us to buy very few conference
proceedings (mainly from the rather economic {\it ASP Conf.\ Series}\,)
and IAU Symposia. The budget for books and other proceedings
depends on the allocation of special funds from the Federal
Secretary of Public Education (SEP) and on personal research
projects (usually from CONACyT), and fluctuates between 0 and 
8000~USD/yr. Delays in either the invoices
of journal publishers, or in the payment by our central
library, usually cause an interruption of our subscriptions
during the first few months of each year (Does this sound 
familiar to you\,...?).
Altogether, we are far from the ideal situation described
in my earlier wishlist (Andernach 1998).

\section{Donations}
Thanks to frequent offers of duplicate items from professional
astronomy librarians (e.g.\ v\'{\i}a ASTROLIB) we acquired
an impressive amount of (mostly older) journals and monographs.
Naturally many of these items are of interest to either the
bibliophile or historians of astronomy and physics. While filling
me with pride, it is a shame to store these books some 200~m away
from our offices where hardly any of us finds the leisure to go
and browse the shelves.

\section{Inventory}
I maintain the inventory as a single ASCII file (of $\sim$300\,kb),
readable by all members of the DA. It saves me to learn dedicated library
software and allows an easy search of items using Unix's {\tt grep} 
command.  For journals I use {\tt bibcode}-style, and free format for the
``rest'' (monographs, theses, manuals, etc., currently $\sim$700 items).
A small excerpt follows: \\[-2.ex]
  
\noindent
\rule{\textwidth}{.3mm}
\noindent
{\scriptsize\tt
1834MmRAS  ~7 + \hspace*{4cm}| 1879MNRAS ~39 - \#2 Suppl. (p.489-560)\\[-.3ex]
1843MmRAS  13 + \hspace*{4cm}| 1891MNRAS ~52 - \#2 Suppl. (p.~67-121)\\[-.3ex]
1843MmRAS  14 + \hspace*{4cm}| 1933MNRAS ~93 - 4-6,8,9 \\[-.3ex]
1847MmRAS  17 + \hspace*{4cm}| 1942MNRAS 101 - 8 only \\[-.3ex]
1849MmRAS  19 + \hspace*{4cm}| 1942MNRAS 102 + \\[-.3ex]
 ...\\[-.3ex]
1879MNRAS  39 - no. 2 Suppl. (p.489-560)\\[-.3ex]
1891MNRAS  52 - no. 2 Suppl. (p.~67-121)\\[1.3ex]
Annals of the Cape Observatory VI, Darling \& Son, 1897.\\[1.3ex]
P.S.\,Barrera,\,E.\,Castro,\,J.R.\,Garza,\,J.J.\,Martinez,\,R.\,Aguirre:\,Memorias\,del\,Gran\,Eclipse\,del\,Sol, \\[-.3ex]
Montemorelos, Nuevo Le\'on, 
28 mayo 1900, Universidad Aut\'onoma de Nuevo Le\'on, 114\,pp. \\[1.3ex]
M.\,de\,Broglie:\,X-rays,\,Transl.~by J.R. Clarke, E.P.\,Dutton \& Co.\,Publishers, 
1922, 204 pp. \\[1.3ex]
D.\,S.\,De\,Young:\,The physics of extragalactic\,radio\,sources,\,Univ.\,Chicago\,Press,\,2002,\,558 pp. \\[-.4ex]
}
\rule{\textwidth}{.3mm}

\normalsize
\section{Other Places in Mexico}

There are now eight places in Mexico where professional astronomers 
work (Phillips et al., 2002). A rough map of their location is
shown in Fig.~1. Only IA-UNAM, INAOE, and OAN (\#1--3) have
long histories and stable budgets for \linebreak[4]

\begin{figure}[hb]
\plotfiddle{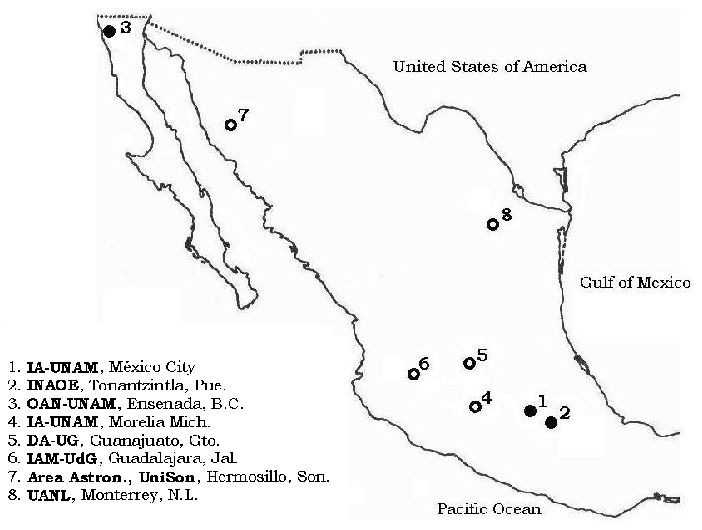}{63mm}{0}{120}{120}{-210}{-130}
\caption{Current centres of professional astronomy in Mexico.}
\end{figure}

\noindent
library and librarian.  All others (\#\,4\,--\,8) were 
established during the last decade. 
The following table gives an overview of the library situation 
at these five ``new'' places. The last column gives a comparison 
with the OAN library at Ensenada, maintained by UNAM.
Numbers in brackets are either uncertain or very variable.
The budget is listed for journals $+$ books. 


\begin{center}
\small
\narrower
\begin{tabular}{|l||c|c|c|c|c||c|}
\hline
  ~Location and         & {\footnotesize Morelia}   & {\footnotesize DAUG} & {\footnotesize IAM-} & U.\,Son- & {\footnotesize Monter-}  &  {\footnotesize Ensen-} \\
\hline
  ~ID\,\# in Fig.\,1    &    (4)  &  (5) & {\footnotesize UdG}~(6) & {\footnotesize ora}~(7)       & {\footnotesize rey}~(8)       & {\footnotesize ada}~(3) \\
\hline \hline
Research Staff & 19 & 8 & 5 & 4 & 1  &  25 \\
\hline
Libarian(s) available  & - & - & (2) & - & (1) & 1 \\
\hline
Library Space exists\,?  &  yes  &  (yes)  &  no    & (yes)  &  no &  yes  \\
\hline
\# journals subscribed &   7  &  10  &  7  &  -- &  8  &  62 \\
\hline
Subscription rate &  inst.  &  pers.  &  inst.  &  -- & pers.  &  inst. \\
\hline
\# back years available  &  $\sim$20  & $\sim$50  &  2.5  &  (2)  &  6--20  & $\sim$20 \\
\hline
budget/year~[1K\,USD]  &  10+4  &  3+(2)  &   3.5+?  &  -- &  $\le$\,1 &  33+15  \\
\hline
Inventory exists? &   no   &   yes  &  no   &  (yes)  &  (yes)  &   yes  \\
\hline
\end{tabular}
\end{center}

\noindent
Astronomy acquisitions for IAM-UdG and Monterrey are made by 
their central libraries, causing a lack of transparency and 
communication between these and the astronomers. At DAUG we have 
no budget for binding journals. However, in 2001 we used some 
left-over monies for binding a few years of {\it ApJ} to find out
that it was $\sim$\,3 times cheaper to bind them in Mexico City than in Guanajuato.

\section{Conclusion}
The changes in the job market and the Internet have affected
radically not only the way astronomers work, but also how an
astronomy library is run, especially at small and ``poor'' places.
Today small groups of astronomers are established independently of
favorable sky conditions and rely mainly on an adequate Internet
connection, but often have to work without a professional librarian.
While this may work ``well'', i.e.\ with little effect on research
output as in our case, it certainly relies heavily on the services
provided by a few professional librarians thinking far beyond
their own institution. I see a dangerous trend for a future
2-class system of astronomy institutions: those with professional
librarians working almost ``behind the scenes'' and those which
have to survive without a local librarian altogether. \\[-1ex]

\acknowledgments
P.\ Phillips,~ L.F.\ Rodr\'{\i}guez,~ A.\ S\'anchez-Ibarra, \linebreak[4]
P.\ Vald\'es Sada, and M.-E.\,Jim\'enez 
kindly provided data on their research centres. Thanks to 
U.\,Grothkopf and S.\,Stevens-Rayburn for comments, to
A.\,Roy and N.\,Loiseau for printing the poster, and to 
L.\,Hdz.~Mendieta for help with Figure\,1.
My attendance of LISA\,IV was financed by CONACyT grant E-27602.

\end{document}